\documentstyle[aps,manuscript]{revtex}

\draft

\begin{document}
\title{Stochastic approach to inflation II: \\
classicality, coarse-graining and noises.}
\author{H. Casini, R. Montemayor}
\address{Instituto Balseiro and Centro At\'{o}mico Bariloche\\
Universidad Nacional de Cuyo - CNEA \\
E. Bustillo 9500, (8400) San Carlos de Bariloche, R\'{\i}o Negro,\\
Argentina.}
\author{P. Sisterna}
\address{Departamento de F\'{\i }sica, Facultad de Ciencias Exactas y
Naturales \\
Universidad Nacional de Mar del Plata \\
Funes 3350, (7600) Mar del Plata, Buenos Aires, Argentina. }
\maketitle

\begin{abstract}
In this work we generalize a previously developed semiclassical approach to
inflation, devoted to the analysis of the effective dynamics of
coarse-grained fields, which are essential to the stochastic approach to
inflation. We consider general non-trivial momentum distributions when
defining these fields. The use of smooth cutoffs in momentum space avoids
highly singular quantum noise correlations and allows us to consider the
whole quantum noise sector when analyzing the conditions for the validity of
an effective classical dynamical description of the coarse-grained field. We
show that the weighting of modes has physical consequences, and thus cannot
be considered as a mere mathematical artifact. In particular we discuss the
exponential inflationary scenario and show that colored noises appear with
cutoff dependent amplitudes.
\end{abstract}


\section{Introduction}

Inflation solves several difficulties that arise in the very early universe,
such as the horizon, flatness and monopole problems, and besides this it
provides a mechanism for the creation of primordial density fluctuations
needed to explain the structures which are now present in the Universe\cite
{starobinsky,linde90}. The most widely accepted approach assumes that the
inflationary stage is driven by a quantum scalar field $\varphi $ with a
potential $V(\varphi )$. Within this perspective, stochastic inflation seeks
to describe the dynamics of this quantum field on the basis of a splitting
of $\varphi $ in a homogeneous and an inhomogeneous component. Usually the
homogeneous one is interpreted as a classical field that arises from a
coarse-grained average over a volume larger than the observable universe,
and plays the role of a global order parameter\cite{goncharov}. All
information on scales smaller than this volume, such as density
fluctuations, is contained in the inhomogeneous component. Although this
theory is widely used and accepted as a general framework, it presents
inconsistencies and has been subjected to several criticisms\cite
{hm,habib,matacz}. Its main problems are related to the treatment of the
global order parameter as a classical field, and the description of the
quantum fluctuations as classical ones.

In a previous work\cite{bcms} we assumed that the coarse-graining volume is
defined so that it leads to a Heaviside function in momentum space. This is
a choice that simplifies the mathematical development because it leads to a
noise with a very simple white spectrum, but at the same time with singular
correlations. A usual assertion is that the dynamics for the coarse-grained
field is not very sensitive to this choice, but its actual implications on
the resulting dynamics are not well understood. The present paper is mainly
devoted to the study of this last point, together with an analysis of
classicality conditions (in our case commutativity conditions). To do this
we develop a general formalism for coarse-graining volumes with arbitrary
shape in momentum space, and analyze the consequences of this shape on the
emergence and structure of the classical effective regime. Besides this
there is another improvement respect to our previous work. There we stated a
sufficient condition for a classical description, neglecting the
contribution of a sector of the noise\cite{old}. Here we analyze the whole
quantum noise sector, and thus the characterization of our classicality
criterion is complete. This analysis enhances our understanding of the role
of the shape of the coarse-graining in the effective classical dynamics, and
also provides us with a well defined regularization scheme to treat the
usual sharp cutoff in momentum space.

In the work just mentioned\cite{bcms} we analyze the emergence of a
classical behavior of the order parameter on the basis of a semiclassical
approach. The inflaton field Lagrangian is:
\begin{equation}
{\cal L}(\varphi ,\varphi _{,\mu })=-\sqrt{-g}\left[ \frac{1}{2}\left(
g^{\mu \nu }\varphi _{,\mu }\varphi _{,\nu }\right) +V(\varphi )\right]
=a^{3}\left( \frac{1}{2}\dot{\varphi}^{2}-\frac{1}{2a^{2}}(\nabla \varphi
)^{2}-V(\varphi )\right) \,,
\end{equation}
for a Friedman-Robertson-Walker metric, $ds^{2}=-dt^{2}+a(t)^{2}\ d\vec{r}
^{2}$. The equation of motion that results for the scalar field operator is
\begin{equation}
\ddot{\varphi}-\frac{1}{a^{2}}\nabla ^{2}\varphi +3H\dot{\varphi}+V^{\prime
}(\varphi )=0\,,
\end{equation}
where the overdot represents the time derivative and $V^{\prime }(\varphi )=
{\frac{dV}{d\varphi }}$, and the metric, given by the Hubble parameter $H=
{\frac{\dot{a}}{a}}$, evolves according to:
\begin{equation}
H^{2}=\frac{4\pi }{3M_{p}^{2}}<\dot{\varphi}^{2}+\frac{1}{a^{2}}(\vec{\nabla}
\varphi )^{2}+2V(\varphi )>\,.
\end{equation}
We decompose the scalar field in its mean value, which by assumption
satisfies a classical equation of motion, plus the quantum fluctuations,
$\varphi =\phi_{cl}+\phi$ with $<\phi >=0$, up to linear terms in $\phi$. In
such a way the equations of motion reduce to a set of two classical
equations which give the evolution of the field $\phi _{cl}$ and the Hubble
parameter. To be consistent with the FRW metrics, we assume that $\phi _{cl}$
is a homogeneous field, and thus we have the following classical equations:
\begin{eqnarray}
&&\ddot{\phi}_{cl}+3H\dot{\phi}_{cl}+V^{\prime }(\phi _{cl})=0,  \label{4} \\
&&H^{2}=\frac{8\pi ^{2}}{3M_{p}^{2}}\rho \,,  \label{5}
\end{eqnarray}
where $\rho $ is the energy density, $\rho =\frac{1}{2}\dot{\phi}_{cl}^{2}
+V(\phi _{cl})$, and one operatorial equation for the quantum fluctuations:
\begin{equation}
\ddot{\phi}-\frac{1}{a^{2}}\nabla ^{2}\phi +3H\dot{\phi}+V"(\phi _{cl})\phi
=0\,.  \label{6}
\end{equation}
In this last equation $H$ and $V_{cl}"$ are functions of $t$ given by Eqs.
(\ref{4}-\ref{5}). In this context we developed the analysis of the emergence
of a classical regime for the inflationary dynamics\cite{bcms}.

The characteristic timescale for the inflaton field can be defined by $\tau
_{d}=\frac{\phi _{cl}}{\dot{\phi}_{cl}},$ and hence its relation with the
Hubble timescale $\tau _{H}=H^{-1}$ is given by:
\begin{equation}
\vartheta \equiv \frac{\tau _{d}}{\tau _{H}}=\sqrt{\frac{2}{3}}\frac{2\pi }
{M_{p}}\frac{\phi _{cl}}{\dot{\phi}_{cl}}\rho ^{1/2}\,,
\end{equation}
and the number of e-folds in a given period of time is:
\begin{equation}
N_{c}=\int_{t_{0}}^{t_{0}+\delta t}dt\;H=\int_{\phi _{0}}^{\phi _{cl}}d\phi
_{cl}^{\prime }\;\frac{\vartheta }{\phi _{cl}^{\prime }}\,.
\end{equation}

If we are interested in an exponential inflationary period, i.e. when the
slow roll of the field holds, then the conditions $\Theta =\frac{M_{p}^{2}}
{4\pi }\left( \frac{H^{\prime }}{H}\right) ^{2}\ll 1$ and $\frac{M_{p}^{2}}
{4\pi }\frac{H^{\prime \prime }}{H}\ll 1$ must be satisfied\cite{copeland}.
The end of inflation, when the scale factor stops accelerating, is given
precisely by $\Theta (\phi _{cl})=1$, which determines $\phi _{cl}^{end}$.
At this point we have $\dot{\phi}_{cl}^{end}\simeq -\frac{V^{\prime }(\phi
_{cl})}{3H}$ and $H^{2}=\frac{8\pi ^{2}}{3M_{p}^{2}}V(\phi _{cl})$, so that:
\begin{equation}
\vartheta \sim -\frac{8\pi ^{2}\phi _{cl}}{M_{p}^{2}}\frac{V(\phi _{cl})}
{V^{\prime }(\phi _{cl})}\,,
\end{equation}
and
\begin{equation}
N_{c}=-\frac{8\pi ^{2}}{M_{p}^{2}}\int_{\phi _{0}}^{\phi _{cl}}d\phi
_{cl}^{\prime }\;\frac{V(\phi _{cl}^{\prime })}{V^{\prime }(\phi
_{cl}^{\prime })}\,.
\end{equation}
A solution to the horizon problem requires $N_{c}\gtrsim 60$, and this in
general implies that $\tau _{d}>\tau _{H}$.

The study of the quantum component becomes much simpler if we redefine the
$\phi$ field such that the equation of motion (\ref{6}) does not have a first
order term, $\phi =e^{-\frac{3}{2}\int dt\ H}\chi $. The equation of motion
for the field operator $\chi $ is:
\begin{equation}
\ddot{\chi}-\frac{1}{a^{2}}\nabla ^{2}\chi -\frac{k_{0}^{2}}{a^{2}}\chi =0\,,
\label{11}
\end{equation}
where $k_{0}^{2}=a^{2}\left( \frac{9}{4}H^{2}+\frac{3}{2}\dot{H}
-V"_{c}\right) $. Thus $\chi $ can be interpreted as a free scalar field
with a time dependent mass parameter. It can be expanded in a set of modes
$\xi _{k}(t)e^{i\vec{k}.\vec{r}}$:
\begin{equation}
\chi (\vec{r},t)=\frac{1}{(2\pi )^{3/2}}\int d^{3}k\left[ a_{k}\xi
_{k}(t)e^{i\vec{k}.\vec{r}}+hc\right] \,,
\end{equation}
where the annihilation and creation operators satisfy the usual commutation
relations for bosons:
\begin{equation}
\lbrack a_{k},a_{k^{\prime }}^{\dagger }]=\delta (\vec{k}-\vec{k}^{\prime
})\qquad ,\qquad [a_{k},a_{k^{\prime }}]=[a_{k}^{\dagger },a_{k^{\prime
}}^{\dagger }]=0\,,
\end{equation}
while the modes are defined for the equation of motion
\begin{equation}
\ddot{\xi}_{k}+\omega _{k}^{2}\xi _{k}=0\,,  \label{14}
\end{equation}
with $\omega _{k}^{2}=a^{-2}\left( k^{2}-k_{0}^{2}\right) $. The function
$k_{0}^{2}(t)$ gives the threshold between an unstable infrared sector
($k^{2}<k_{0}^{2}$), which includes only long wavelengths relative to the
coarse-graining scale, and a stable short wavelength sector
($k^{2}>k_{0}^{2} $). We adopt the normalization condition $\xi _{k}\dot{\xi}
_{k}^{*} -\dot{\xi}_{k}\xi _{k}^{*}=i$ for the modes, such that the field
operators $\chi $ and $\dot{\chi}$ satisfy canonical commutation relations.

In the section below we introduce a weight function to define the
coarse-grained field and develop the general formalism, showing the
importance of an adequate definition of the quantum noises and the
conditions to properly derive a classical regime. After that, in Section
III, we apply this approach to the inflationary inflation scenario and show
the need for considering the proposed noise definition to prove the
emergence of a classical regime. The last section is devoted to some
concluding remarks.

\section{The general approach}

In order to define the coarse-grained field, an effective field which is an
average of the long wavelength modes, we use a weight function. We assume
that this is an isotropic function which contains only one parameter $b$
with length dimensions, so that it is of the form $b^{-3}g(r/b)$, which is
always larger than the causal horizon. The coarse-grained field $\chi _{b}$
is defined by:
\begin{equation}
\chi _{b}\equiv \frac{1}{(2\pi )^{3/2}}\frac{1}{b^{3}}\int d^{3}rg(r/b)\chi
(\vec{r},t)=\frac{1}{(2\pi )^{3/2}}\int d^{3}k\;G(\vec{k})\;\left[ a_{k}\xi
_{k}(t)+hc\right] \, ,  \label{15}
\end{equation}
with:
\begin{equation}
G(\vec{k})=\frac{1}{(2\pi )^{3/2}}\int \frac{d^{3}r}{b^{3}}e^{i\vec{k}.
\vec{r}}g(r/b)=\sqrt{\frac{2}{\pi }}\frac{1}{kb^{3}}\int dr\ r\ g(r/b)
\sin {kr} \, ,
\end{equation}
or reciprocally:
\begin{equation}
g(r/b)=\sqrt{\frac{2}{\pi }}\frac{b^{3}}{r}\int dk\;k\sin kr\;G(k)\, .
\end{equation}
These equations can be written in terms of the dimensionless variables
$\beta =kb$ and $\rho =r/b$ as follows:
\begin{eqnarray}
G(\beta ) &=&\sqrt{\frac{2}{\pi }}\frac{1}{\beta }\int d\rho \ \rho \
\sin {\kappa \rho }\ g(\rho )\, , \\
g(\rho ) &=&\sqrt{\frac{2}{\pi }}\frac{1}{\rho }\int d\beta \;\beta \;\sin
\beta \rho \;G(\beta )\, ,
\end{eqnarray}
from Eq. (\ref{15}), we obtain for the derivatives of $\chi _{b}$:
\begin{eqnarray}
\dot{\chi}_{b} &=&\frac{1}{(2\pi )^{3/2}}\int d^{3}k\left[ \dot{G}(\beta
)a_{k}\xi _{k}(t)+G(\beta )a_{k}\dot{\xi}_{k}(t)+hc\right] \, , \\
\ddot{\chi}_{b} &=&\frac{1}{(2\pi )^{3/2}}\int d^{3}k\left\{ a_{k}\left[
\left( \ddot{G}(\beta )-\omega _{k}^{2}g_{b}\right) \xi _{k}+2\dot{G}(\beta)
\dot{\xi}_{k}\right] +hc\right\} \, .
\end{eqnarray}
The parameter $b$ is chosen so that the $k^2$ term, that is, the noise
independent of the weight function, can be neglected in the equation of
motion for $\chi _{b}$. The appropriate value can be obtained from the
equation of motion for the modes, Eq.(\ref{14}). We can choose the
characteristic length of the distribution larger than the horizon scale,
i.e. $b^{-1}=\varepsilon k_{0}$ with $\varepsilon \ll 1$. From the
definition of $\chi _{b}$ and the expression for $\ddot{\chi}_{b}$, assuming
that we are only considering infrared modes with $k^{2}\ll k_{0}^{2}$, we
obtain the equation of motion for the coarse-grained field\cite{old}:
\begin{equation}
\ddot{\chi}_{b}-\frac{k_{0}^{2}}{a^{2}}\chi _{b}=\eta _{b}+\kappa _{b}\, ,
\label{22}
\end{equation}
where:
\begin{eqnarray}
\eta _{b} &=&\frac{1}{(2\pi )^{3/2}}\int d^{3}k\left[ a_{k}\ddot{G}(\beta)
\xi _{k}+hc\right] \, , \\
\kappa _{b} &=&\frac{2}{(2\pi )^{3/2}}\int d^{3}k\left[ a_{k}\dot{G}(\beta)
\dot{\xi}_{k}+hc\right] \, .
\end{eqnarray}

The dependence of the function $G(\beta )$ on $t$ is given only through the
parameter $b$, and thus, using the relation $\partial _{b}G=\frac{k}{b}
\partial _{k}G$, we can write:
\begin{eqnarray}
\dot{G}(\beta ) &=&\frac{\dot{b}\beta }{b}\partial _{\beta }G(\beta)\, ,
\label{25} \\
\ddot{G}(\beta ) &=&\left( \frac{\ddot{b}\beta }{b}\right) \partial _{\beta
}G(\beta )+\left( \frac{\dot{b}\beta }{b}\right) ^{2}\partial _{\beta
}^{2}G(\beta )\, .  \label{26}
\end{eqnarray}

Taking into account the algebra for the creation and annihilation operators,
the operators $\chi _{b}$, $\eta _{b}$ and $\kappa _{b}$ satisfy the
commutation relations :
\begin{eqnarray}
\lbrack \chi _{b}(t),\eta _{b}(t^{\prime })] &=&0\, ,  \label{27} \\
\lbrack \chi _{b}(t),\kappa _{b}(t^{\prime })] &=&\frac{2}{(2\pi )^{3}}\int
d^{3}kG(\beta )\dot{G}(\beta ^{\prime })\left( \xi _{k}(t)\dot{\xi}_{k}^{*}
(t^{\prime })-\xi _{k}^{*}(t)\dot{\xi}_{k}(t^{\prime })\right) \, , \\
\lbrack \eta _{b}(t),\kappa _{b}(t^{\prime })] &=&\frac{2}{(2\pi )^{3}}\int
d^{3}k\ddot{G}(\beta )\dot{G}(\beta ^{\prime })\left( \xi _{k}(t)\dot{\xi}
_{k}^{*}(t^{\prime })-\xi _{k}^{*}(t)\dot{\xi}_{k}(t^{\prime })\right) \, ,
\label{29}
\end{eqnarray}
and the quantum noises have the correlation functions:
\begin{eqnarray}
&<&\kappa _{b}(t)\kappa _{b}(t^{\prime })>=\frac{4}{(2\pi )^{3}}\int
d^{3}k\; \dot{G}(\beta )\dot{G}(\beta ^{\prime })\left( \dot{\xi}_{k}(t)
\dot{\xi} _{k}^{*}(t^{\prime })\right) \, ,  \label{30} \\
&<&\eta _{b}(t)\eta _{b}(t^{\prime })>=\frac{1}{(2\pi )^{3}}\int d^{3}k\;
\ddot{G}(\beta )\ddot{G}(\beta ^{\prime })\left( \xi _{k}(t)\xi
_{k}^{*}(t^{\prime })\right) \, ,  \label{31} \\
&<&\kappa _{b}(t)\eta _{b}(t^{\prime })>=\frac{2}{(2\pi )^{3}}\int d^{3}k\;
\dot{G}(\beta )\ddot{G}(\beta ^{\prime })\left( \dot{\xi}_{k}(t)\xi
_{k}^{*}(t^{\prime })\right) \, ,  \label{32}
\end{eqnarray}
with $\beta =kb(t)$ and $\beta^{\prime}=kb(t^{\prime })$. The quantum
character of the fields becomes apparent through the non-null commutation
relations and the complex correlation functions of the noises. To have an
effective classical theory it is necessary that the non-null commutators be
irrelevant, and consequently that the correlations be real.

To simplify the discussion we will assume that $G(\beta)$ is non null only
in the range $0<\beta \lesssim \bar{\beta}$, and that its derivatives are
practically null for all values of $\beta $, except in a domain $(\bar{\beta}
-\frac{\Delta \beta }{2},\bar{\beta}+\frac{\Delta \beta }{2})$, with $\Delta
\beta \ll \bar{\beta}$, where the modes $\xi _{k}$ vary slowly. This last
interval corresponds to the ''wall '' of the coarse-graining domain, which
we consider to be relatively well defined. Under these assumptions Eqs.(\ref
{27}-\ref{29}) at $t=t^{\prime }$ can be approximately written:
\begin{eqnarray}
\lbrack \chi _{b}(t),\kappa _{b}(t)] &\simeq &-\frac{3}{\pi ^{2}} \frac{\dot{
b}\bar{\beta}^{3}}{b^{4}}G\left( \bar{\beta}\right) \;\left. \mathop{\rm Im}
\left( \xi _{k}\dot{\xi}_{k}^{*}\right) \right| _{k=\bar{\beta}/b}\, , \\
\lbrack \eta _{b}(t),\kappa _{b}(t)] &\simeq &\frac{4}{\pi ^{2}}\frac{\dot{b}
\ddot{b}\bar{\beta}^{4}}{b^{5}}\int d\beta \;\left( \partial _{\beta
}G\right) ^{2}\;\left. \mathop{\rm Im} \left( \xi _{k}\dot{\xi}
_{k}^{*}\right) \right| _{k=\bar{\beta}/b}\, ,
\end{eqnarray}
where the $b,$ $\xi _{k}$ and $G$ functions are evaluated at $t$. Here it
becomes evident that the main difficulty in considering the equations of
motion as classical is the $\kappa $ operator. It does not commute with
$\chi $ and $\eta$, whereas the latter ones do between them. This drawback
can be overcome only if the contribution of $\kappa $ is negligible compared
with that of $\eta $. The contributions of the different terms can be
weighted by their rms values. Assuming that the assumptions discussed before
hold, we have:
\begin{eqnarray}
<\kappa _{b}(t)\kappa _{b}(t)> &=&\left. \left| \dot{\xi}_{k}\right|
^{2}\right| _{k=\bar{\beta}/b}I_{1}\, , \\
<\eta _{b}(t)\eta _{b}(t)> &=&\left. \left| \xi _{k}\right| ^{2}\right| _{k=
\bar{\beta}/b}I_{2} \, .
\end{eqnarray}
Thus a necessary condition for classicality is:
\begin{equation}
\left| \frac{<\kappa _{b}(t)\kappa _{b}(t)>}{<\eta _{b}(t)\eta _{b}(t)>}
\right| \sim \left. \frac{\left| \dot{\xi}_{k}(t)\right| ^{2}}{\left| \xi
_{k}(t)\right| ^{2}}\right| _{k=\bar{\beta}/b}\left| \frac{I_{1}}{I_{2}}
\right| \ll 1\, ,  \label{38}
\end{equation}
where $I_{1}$ and $I_{2}$ are given by
\begin{eqnarray}
I_{1} &\simeq &\frac{2}{\pi ^{2}}\frac{\dot{b}^{2}\bar{\beta}^{4}}{b^{5}}
\int d\beta \;\left( \partial _{\beta }G\right) ^{2}\, , \\
I_{2} &\simeq &\left( \frac{\ddot{b}}{2\dot{b}}\right) ^{2}I_{1}+ \frac{\dot
{b }^{4}\bar{\beta}^{6}}{b^{7}}\int d\beta \;\left( \partial
_{\beta}^{2}G\right) ^{2}\, ,
\end{eqnarray}
and hence
\begin{equation}
\frac{I_{2}}{I_{1}}\simeq \frac{1}{4}\left( \frac{\dot{b}}{b}\right)
^{2}\left( \left( \frac{\ddot{b}b}{\dot{b}^{2}}\right) ^{2}+2\pi ^{2}\bar{
\beta}^{2}\frac{\int d\beta \;\left( \partial _{\beta }^{2}G\right) ^{2}}{
\int d\beta \;\left( \partial _{\beta }G\right) ^{2}}\right) \gtrsim \frac{1
}{4}\left( \frac{\dot{b}}{b}\right) ^{2} \, .
\end{equation}
From here we can state a necessary condition
\begin{equation}
\left. \frac{\left| \dot{\xi}_{k}(t)\right| }{\left| \xi _{k}(t)\right| }
\right| _{k\simeq \bar{\beta}/b}\ll \frac{\dot{b}}{b}\, ,  \label{43}
\end{equation}
for relation (\ref{38}) to hold. Given the coarse-grained field defined by
$G(\beta)$ and the character of inflation given by $H$, this relation states
a condition to be satisfied by the modes in the threshold sector between the
unstable infrared sector and the stable short wavelength sector
($k^{2}>k_{0}^{2}$), given by $k_{0}=\frac{1}{\varepsilon b}$, so that the
coarse-grained field admits a classical description.

To have a classical regime there is another condition to be satisfied,
namely the correlation function of $\eta (t)$ must be real. In general the
correlation functions decrease rapidly with $(t-t^{\prime })$, and hence we
can approximate $t^{\prime }\simeq t+\delta t$, and take the leading order
term using $\xi_{k}(t^{\prime})\simeq \xi_{k}(t)+\dot{\xi}_{k}(t)\delta t$.
To this time increment correponds a $\beta $ variation $\delta \beta \sim
\frac{\beta \dot{b}}{b}\delta t$. Up to linear contributions in $\delta t$
the $\eta $ correlation function is:
\begin{equation}
<\eta _{b}(t)\eta _{b}(t^{\prime })>=\frac{1}{(2\pi )^{3}}\int d^{3}k\;
\ddot{G}(\beta )\ddot{G}(\beta +\delta \beta )\left| \xi _{k}\right| ^{2}
\left( 1+\frac{\dot{\xi}_{k}^{*}}{\xi _{k}^{*}}\delta t\right) \,.
\label{42}
\end{equation}
Hence the condition to have a real $\eta $ correlation function becomes
\begin{equation}
\mathop{\rm Im}\left. \frac{\dot{\xi}_{k}}{\xi _{k}}\right| _{k=\bar{\beta}
/b}\ll \delta t^{-1}\simeq \frac{\beta }{\delta \beta }\frac{\dot{b}}{b}\ll
\frac{\dot{b}}{b}\,,  \label{44}
\end{equation}
but if (\ref{43}) is satisfied this last relation is also satisfied.
Therefore, in general, when $\eta $ dominates its time correlations are
practically real.

In fact, these conditions are not sufficient to ensure that we have an
effective classical dynamics because, although they warrant that the
correlation $<\kappa _{b}(t)\kappa _{b}(t^{\prime })>$ is negligible with
respect to $<\eta _{b}(t)\eta _{b}(t^{\prime })>$ and also that this last
correlation function can be considered as a real one, they are not enough to
warrant that $<\kappa _{b}\eta _{b}+\eta _{b}\kappa _{b}>$ is simultaneously
negligible.

As was already pointed out in our previous article, the noises appear only
in the form of the right hand side term of Eq.(\ref{22}), so that its
decomposition in terms of $\kappa $, and $\eta $ is not uniquely defined. We
can use this freedom to minimize the weight of the non-commuting operator,
and thus to optimize the effective classical description\cite{bcms}.
Following the preceding paper, we introduce a new partition for the noise
term in the equation of motion, according to:
\begin{eqnarray}
\tilde{\eta}_{b} &=&(1+s)\eta _{b}\, , \\
\tilde{\kappa}_{b} &=&\kappa _{b}-s\eta _{b}\, .
\end{eqnarray}

For these new operators we have:
\begin{eqnarray}
\lbrack \chi _{b}(t),\tilde{\eta}_{b}(t)] &=&(1+s)[\chi _{b}(t),\eta
_{b}(t)]=0\,, \\
\lbrack \chi _{b}(t),\tilde{\kappa}_{b}(t)] &=&[\chi _{b}(t),\kappa
_{b}(t)]\,, \\
\lbrack \tilde{\eta}_{b}(t),\tilde{\kappa}_{b}(t)] &=&(1+s)[\eta
_{b}(t),\kappa _{b}(t)]\,.
\end{eqnarray}
In order to optimize the classical description we minimize the correlations
$<\tilde{\kappa}_{b}\tilde{\kappa}_{b}>$ and $<\tilde{\eta}_{b}\tilde{\kappa}
_{b}>$ . Assuming that the correlations are real we have
\begin{equation}
\left. \frac{<\tilde{\kappa}_{b}(t-\delta t)\tilde{\kappa}_{b}(t+\delta t)>}
{<\tilde{\eta}_{b}(t-\delta t)\tilde{\eta}_{b}(t+\delta t)>}\right|
_{k=k_{o} \bar{\beta}}\ll 1\,.  \label{50}
\end{equation}

The function that minimizes the relation between the expectation values is:
\begin{equation}
s(t)=\frac{<\kappa _{b}\kappa _{b}>+1/2<\kappa _{b}\eta _{b}+\eta _{b}\kappa
_{b}>}{<\eta _{b}\eta _{b}>+1/2<\kappa _{b}\eta _{b}+\eta _{b}\kappa _{b}>}
\,.
\end{equation}
This $s(t)$ also minimizes the relation (\ref{50}) at first order in $\delta
t$. From here we have
\begin{equation}
\frac{<\tilde{\kappa}_{b}\tilde{\kappa}_{b}>}{<\tilde{\eta}_{b}\tilde{\eta}
_{b}>}=\frac{<\kappa _{b}\kappa _{b}><\eta _{b}\eta _{b}>-1/4<\kappa
_{b}\eta _{b}+\eta _{b}\kappa _{b}>^{2}}{(<\kappa _{b}\kappa _{b}>+<\eta
_{b}\eta _{b}>+<\kappa _{b}\eta _{b}+\eta _{b}\kappa _{b}>)<\eta _{b}\eta
_{b}>}\,.
\end{equation}
Furthermore, if we compute the correlation between $\tilde{\kappa}_{b}$ and
$\tilde{\eta}_{b}$ with this value for $s$ we have:
\begin{equation}
\frac{<\tilde{\kappa}_{b}\tilde{\eta}_{b}>}{<\tilde{\eta}_{b}\tilde{\eta}
_{b}>}=-4\frac{<\tilde{\kappa}_{b}\tilde{\kappa}_{b}>}{<\tilde{\eta}_{b}
\tilde{\eta}_{b}>}\,,
\end{equation}
which implies that the same condition makes both of them negligible.
Therefore, the condition for disregarding the contribution of $\tilde{\kappa}
_{b}$, and hence having a valid classical description, is characterized by:
\begin{equation}
Q\equiv \left| \frac{<\kappa _{b}\kappa _{b}><\eta _{b}\eta _{b}>-1/4<\kappa
_{b}\eta _{b}+\eta _{b}\kappa _{b}>^{2}}{(<\kappa _{b}\kappa _{b}>+<\eta
_{b}\eta _{b}>+<\kappa _{b}\eta _{b}+\eta _{b}\kappa _{b}>)<\eta _{b}\eta
_{b}>}\right| \ll 1\,.  \label{54}
\end{equation}

\section{Exponential inflationary scenario}

We will now apply the preceding approach to the exponential inflationary
scenario. If the system is at a minimum $V_{0}$ of the instanton potential
$V(\phi )$, the classical solution $\varphi _{cl}$ is a constant field and
the fluctuations become a quantum field with a mass $m^{2}=\left. \frac
{d^{2}V}{d\varphi ^{2}}\right| _{\varphi _{cl}}$ and the Hubble constant is
$H=\sqrt{\frac{4\pi V_{0}}{3M_{p}^{2}}}$. Therefore, the scale factor is
$a(t)=e^{Ht}$ and the threshold parameter is given by $k_{0}=\nu He^{Ht}$,
where $\nu =\sqrt{\frac{9}{4}-\frac{m^{2}}{H^{2}}}$. Thus the equation of
motion for the modes is:
\begin{equation}
\ddot{\xi}_{k}+(k^{2}e^{-2Ht}-\nu ^{2}H^{2})\xi _{k}=0\,,
\end{equation}
and its general solution can be written:
\begin{equation}
\xi _{k}(t)=A_{1}H_{\nu }^{(1)}\left( \frac{k}{H}e^{-Ht}\right) +A_{2}H_{\nu
}^{(2)}\left( \frac{k}{H}e^{-Ht}\right) \,.
\end{equation}
We will use the boundary conditions which correspond to the Bunch-Davies
\cite{bunch-davies} vacuum, leading to
\begin{equation}
\xi _{k}(t)=\frac{1}{2}\sqrt{\frac{\pi }{H}}H_{\nu }^{(2)}\left( \frac{k}{H}
e^{-Ht}\right) \,.  \label{56}
\end{equation}
This is a complex wave function. To satisfy a classical interpretation the
correlation of the noise must be real. The real part of the wave function
(\ref{56}) decreases exponentially with $t$, and in this case it is
responsible for the imaginary part of the noise correlation. In our case
$0<k\lesssim \varepsilon k_{0}$, and thus we can state the condition to have
a real noise correlation as $e^{-Ht}\ll H/\left( \varepsilon k_{0}\right) $.
This implies that after a long enough time, $t\gg H^{-1}\ln \left( H/\left(
\varepsilon k_{0}\right) \right) $, the wave funcion $\xi _{k}(t)$ can be
considered imaginary and the noise correlations real. The inflationary stage
lasts $N_{c}\gtrsim 60$ number of e-folds, and thus we have a wide margin to
reach real noise correlations. In this case we can use the approximate
expression for the modes:
\begin{equation}
\xi _{k}(t)\simeq -\frac{i}{2}\sqrt{\frac{1}{\pi H}}\Gamma _{(\nu )}\left(
\frac{k}{2H}\right) ^{-\nu }e^{\nu Ht}\,.  \label{57}
\end{equation}

To be specific, we will define the coarse graining average by using a smooth
approximant of the Heaviside function as our weight function:
\begin{equation}
G_{\alpha }(k)=\frac{1}{2}\left[ 1-\mathop{\rm erf}\left( \frac{bk-1}{\alpha
}\right) \right] \,,
\end{equation}
where $\alpha $ parametrizes the family of functions. When $\alpha
\rightarrow 0$ we have $G_{\alpha }(k)\rightarrow \theta (1-bk)$. In this
case, from Eqs.(\ref{25}-\ref{26}) and using the relations $\frac{\dot{b}}{b}
=H$ and $\frac{\ddot{b}}{b}=H^{2}$, we have:
\begin{eqnarray}
\dot{G} &=&-\frac{bHk}{\alpha \sqrt{\pi }}e^{-\left( \frac{bk-1}{\alpha }
\right) ^{2}}\, , \\
\ddot{G} &=&-\frac{bH^{2}k}{\alpha \sqrt{\pi }}\left[ 1-\frac{2bk}{\alpha
^{2}}\left( bk-1\right) \right] e^{-\left( \frac{bk-1}{\alpha }\right)
^{2}}\,,
\end{eqnarray}
where $b\propto e^{Ht}$. Here $\bar{\beta}\approx 1$, and thus the variation
of $t$ is bounded by $\Delta t\sim \frac{\delta \beta }{H}$.

By replacing these expressions and the corresponding derivatives of the mode
$\xi _{k}(t)$, defined by Eq. (\ref{57}), in Eqs.(\ref{30}-\ref{32}), we can
compute the correlation functions for the noises. They depend on integrals
of the form $I_{s}=\int_{0}^{\infty }dkk^{s}e^{-\frac{b^{2}+b^{\prime 2}}
{\alpha^{2}}\left( k-\frac{b+b^{\prime }}{b^{2}+b^{\prime 2}}\right) ^{2}}$,
where $b=b(t)$ and $b^{\prime }=b(t^{\prime })$. When $\alpha $ is small
they are given by the simple expression $I_{s}\approx \alpha \sqrt{\frac{\pi
}{b^{2}+b^{\prime 2}}}\left( \frac{b+b^{\prime }}{b^{2}+b^{\prime 2}}\right)
^{s}$, where we have neglected $\alpha ^{-1}$ exponentially decreasing
terms, and the correlation functions acquire the form:
\begin{eqnarray}
&<&\kappa _{b}(t)\kappa _{b}(t^{\prime })>=4\nu ^{2}W(t,t^{\prime }) \\
&<&\eta _{b}(t)\eta _{b}(t^{\prime })>=\left( 1-\allowbreak \frac{4\left(
bb^{\prime }\right) ^{2}\left( b^{2}-b^{\prime 2}\right) ^{2}}{\alpha
^{4}\left( b^{2}+b^{\prime 2}\right) ^{4}}\right) W(t,t^{\prime })
\label{62} \\
&<&\kappa _{b}(t)\eta _{b}(t^{\prime })>=2\nu \left( 1+\frac{2bb^{\prime }}
{\alpha ^{2}}\frac{b^{2}-b^{\prime 2}}{\left( b^{2}+b^{\prime 2}\right)^{2}}
\right) W(t,t^{\prime })
\end{eqnarray}
where $W(t,t^{\prime })=\frac{2^{2\nu -5}H^{2\nu -3}\Gamma _{(\nu
)}^{2}bb^{\prime }}{\alpha \pi ^{7/2}\sqrt{b^{2}+b^{\prime 2}}}\left(
\frac{b+b^{\prime }}{b^{2}+b^{\prime 2}}\right) ^{4-2\nu }e^{-\frac{\left(
b-b^{\prime }\right) ^{2}}{\alpha ^{2}\left( b^{2}+b^{\prime 2}\right) }+\nu
H\left( t+t^{\prime }\right) }$. Therefore, for $t=t^{\prime }$ we have
\begin{equation}
\frac{<\kappa _{b}(t)\kappa _{b}(t)>}{<\eta _{b}(t)\eta _{b}(t)>}=4\nu
^{2}=\left( 9-\frac{4m^{2}}{H^{2}}\right) \,,
\end{equation}
and thus it seems that only an inflaton with a mass of the order of the
Planck mass, $m\simeq \frac{3}{2}H$, could develop a classical regime for
the coarse grained field. However, if we work with the redefined noise
partition introduced in Section II, the parameter $s(t)$ which optimizes the
classical description is in this approximation:
\begin{equation}
s(t)=2\nu =\sqrt{9-\frac{4m^{2}}{H^{2}}}\,.
\end{equation}
Hence the condition (\ref{54}) for neglecting the contribution of $\tilde
{\kappa}_{b}$ is automatically satisfied for every value of the inflaton mass
because $Q$ $\simeq 0$. Therefore our necessary conditions to have an
effective classical regime are satisfied with a noise given by $\tilde{\eta}
_{b}=(1+2\nu )\eta _{b}$.

Here we have discussed the case of an inflaton field with a non-null mass.
The zero mass case deserves a special discussion. As it is well known, in
this case the modes (\ref{57}) lead to infrared divergences for the
correlation functions of the scalar field, but this does not affect the
correlation functions of the quantum noises, because, according to Eqs.(\ref
{30}-\ref{32}), they have kernels that contain derivatives of the weight
function $G(k)$, which annihilate the contributions of the long-wave modes.
In other words, the noises only live in the walls of the coarse-grained
domains. Of course, the massless limit is not a realistic case and does not
correspond to a truly inflationary process. A massless field driving
inflation must have a nonlinear dynamics, and the nonlinear couplings will
produce a feedback between the quantum fluctuations and the background. In
this case the fluctuations can not be considered small perturbations to the
classical field, which complicates the analysis in a highly non-trivial
way. In general, our approach can be applied to a wide range of cases,
provided that infrared divergencies do not appear\cite{sahni}.

\section{ Final remarks}

This work revises and generalizes our previous semiclassical approach to
inflation, by considering a coarse-grained field with a smooth ''wall '' in
momentum space. One of the most interesting advantages of this approach is
that it allows us to study classicality conditions considering the whole
quantum noise sector for the coarse-grained field, whereas this is not
possible for a sharp cutoff because of the highly singular structure it
produces for the quantum noise correlations. We divide this sector in two: a
noise that commutes at equal times with the coarse-grained field and the
remaining non-commuting noises. The classicality conditions discussed here
arise from two considerations. One involves the relation between the
different quantum noises, and the requirements for the noncommuting sector
to be negligible. The other is related to the imaginary part of the
correlation function of the quantum noise commuting with the coarse-grained
field, which is proportional to the commutator of this quantum noise at
different times. When the different-time commutator can be considered null
the noise and its velocity commute, and thus the noise and its associated
momentum can be considered as c-numbers, which may be viewed as a signal of
classicality.

The redefinition of the noises allows us to obtain a more reliable condition
for the non commuting fluctuations to be negligible. Furthermore, this can
be essential in order to determine the existence of an effective classical
regime, as it is clear in the massless exponential inflation model
considered above. Without this redefinition we showed that the possibility
of a classical regime is not evident, but once this redefinition is
implemented and we turn to the relevant noises the situation changes
completely. It become clear that at early times we do not have classicality
because we cannot neglect either the non-commuting sector or the imaginary
parts of the fluctuations. However, at later times, not only can we neglect
the non commuting sector at equal times, but we can also neglect the time
correlations, and thus we can consider that a classical regime is achieved.

Furthermore, this analysis sheds light on the role of the shape of the wall
of the coarse-grained domain. In the example discussed above, once we assume
that the weight function satisfies (\ref{44}), its shape does not affect the
conditions to have an effective classical regime, although the amplitude and
the spectral characteristics of the resulting classical noise is sensible to
this shape. The amplitude depends on the $\alpha $ parameter in such a way
that it increases when $\alpha $ decreases, diverging when the wall of the
domain is given by a step function, i.e. when $\alpha $ becomes null. The
correlation function (\ref{62}) states that in general the noise is colored,
and not white as generally assumed. Usually the definition of the
coarse-grained field is considered as a mere mathematical artifact, but here
we see that relevant physical features of the model depend on it. This opens
a very interesting question regarding the possible physical origin of the
structure of the domain characterizing the effective field.

Defining a classical behavior for an effective degree of freedom as we do in
this work is significant, but this discussion does not exhaust the problem.
The relevance of an effective degree of freedom is not only dictated by its
classical or quantum behavior, but also by the correlations with the
observables that we are able to measure. In such a sense, our analysis
complements other approaches, such as theories for cosmological
perturbations where not only the matter fields but also the metric are
quantized\cite{brandenberger}, and theories where the classical world
appears through the decoherence of relevant degrees of freedom under the
influence of noise from a non observed sector \cite{hu,paz,zurek}. These
last approaches are mainly based on the path integral techniques and
Feynman-Vernon influence functionals, in which irrelevant degrees of freedom
are integrated out. In the case we consider, the relevant degree of freedom
involves an infinite number of modes, with time dependent weights. Work is
under way dealing with the treatment of these effective degrees of freedom
from the path integral point of view \cite{ckms}.

\section{Acknowledgments}

This research has been partially supported by CONICET, Argentina.


\begin{references}
\bibitem{old}  There is a difference with the constraint obtained in Ref.
\cite{bcms}: there we have a necessary condition by comparing only the first
term of $\eta _{b}$ with $\kappa _{b}$, and not considering the second one
(there named $\lambda )$. In the present treatment we consider the complete
quantum fluctuation.
\end{references}
\end{document}